\documentclass[preprintnumbers,prd,twocolumn,aps,showpacs,amsmath,amssymb,floats]{revtex4-1}
\usepackage{graphicx}
\usepackage{epsfig}
\usepackage{bm}
\usepackage{amsfonts}
\usepackage{amssymb}
\usepackage{subfloat,subfig}
\usepackage{times}
\usepackage{natbib}
\usepackage[titletoc,title]{appendix}
\usepackage{textcomp}
\usepackage{natbib}
\usepackage[normalem]{ulem}

\usepackage[dvips]{color}

\def\lapp{\mathrel{\rlap{\raise.5ex\hbox{$<$}}
                    {\lower.5ex\hbox{$\sim$}}}}
\def\gapp{\mathrel{\rlap{\raise.5ex\hbox{$>$}}
                    {\lower.5ex\hbox{$\sim$}}}}

\begin{document}
\title{Helium reionization in the presence of self-annihilating clumpy dark matter}

\author{Bidisha Bandyopadhyay}
\email{bidisha@physics.du.ac.in}
\affiliation{Department of Physics and Astrophysics, University of Delhi, Delhi, India}
\author{Dominik R.G. Schleicher}
\email{dschleicher@astro-udec.cl}
\affiliation{Departamento de Astronom\'ia, Facultad Ciencias F\'isicas y Matem\'iicas, Universidad de Concepci\'on, Av. Esteban Iturra s/n Barrio Universitario, Casilla 160-C, Concepci\'on, Chile}

\begin{abstract}
The reionization of helium describes the transition from its singly ionized state to a doubly-ionized state in the intergalactic medium (IGM). This process is important for the thermal evolution of the IGM and influences the mean free path of photons with energies above $54.4$~eV. While it is well-known that helium reionization is mostly driven by the contribution of energetic quasars at $z<6$, we study here how helium reionization proceeds if there is an additional contribution due to the annihilation of dark matter. We explore the effects of different dark matter profiles for the dark matter clumping factor, which can significantly enhance the annihilation rate at late times. We find that the presence of dark matter annihilation enhances the He$^{++}$ abundance at early stages where it would be zero within the standard model, and it can further increase during structure formation, reflecting the increase of the dark matter clumping factor. The latter is, 
however, degenerate 
with the build-up of the quasar contribution, and we therefore expect no significant changes at late times. We expect that future studies of the He$^+$ Lyman $\alpha$ forest may help to assess whether the evolution is consistent with the contribution from quasars alone, or if an additional component may be required.
\end{abstract}

\maketitle
\section{Introduction}

About $24\%$  of the baryonic mass in the Universe  consists of primordial helium, implying that the ionization state of helium plays an important role in the thermal evolution of the intergalactic medium (IGM) which influences the mean free path of photons with energies greater than $54.4$ eV. Measurements of the He$^+$ Ly$\alpha$ optical depth suggest that helium has been reionized and now occurs predominantly in the double-ionized state. These measurements are based on Hubble Space Telescope (HST) observations of Q0302-003 \cite{Jakobsen,Heap}, observations from the Hopkins Ultraviolet Telescope and the Far Ultraviolet Spectroscopic Explorer (FUSE) of HS 1700+6416 \cite{Davidsen,Fechner}, HST and FUSE observations of HE 2347−4342 \cite{Reimers1997,Kriss,Smette,Shull}, and HST observations of PKS 1935-692 \cite{Anderson}, SDSS J2346-0016 \cite{ZhengCh,ZhengKr,ZhengMe}, Q1157+3143 \cite{Reimers2005} and SDSS J1711+6052 \cite{ZhengMe}, spanning a redshift range $2.2 < z < 3.8$. 

The reionization of helium (He$^{+}\rightarrow$ He$^{++}$) requires highly energetic 
photons with energies above at least $54.4$~eV, and is therefore accompanied by a significant increase in the thermal energy of the IGM \cite{Haardt,Furlanetto2008}. Such changes in the thermal evolution can have a significant impact on structure formation, in particular in the context of the formation of lower-mass objects like dwarf galaxies \cite{Pfrommer}. Understanding the evolution of helium reionization is therefore important to understand the evolution of the IGM as well as the formation of galaxies within the IGM.\\

Observations from galaxy surveys and CMB signals suggest that most of the matter content of the universe is non-baryonic in nature. In particular, there is a pressureless component representing about $31\%$ of our Universe which is called dark matter (from the PLANCK survey \cite{Planck}), which leaves an imprint on various astrophysical and cosmological observations only due to its gravitational interaction. Many particle physics models suggest that the dark matter consists of weakly interacting massive particles (WIMPs) with typical masses of $10-1000$ GeV which may annihilate into photons, electrons or positrons and neutrinos. 

We assume here an approximately equal distribution among the potential annihilation products, as expected from generic dark matter models. We however note that an annihilation signal would not be visible if the annihilation products consist of neutrinos only. As a result, an  excess of cosmic ray signals from the center of galaxies is  expected due to the self-annihilation of these dark matter particles. From recent studies dark matter masses $M_{\chi}<26$ GeV are ruled out for a thermally averaged annihilation cross-section $\langle\sigma_{\chi} v\rangle = 3.0 \times 10^{-26}  \mathrm{cm}^3 \mathrm{s}^{-1}$ and perfect absorbtion or for a more realistic absorbtion $M_{\chi} < 5$ GeV \cite{Madhavacheril}. The dark matter candidates which are massive and have a negligible free-streaming length compared to the first protogalaxies are considered to be 'cold' while the warm dark matter particles (WDM) are comparably less massive and their  free-streaming length is comparable to the size of protogalaxies.  WDM is a possible alternative to CDM since it can explain the presence of cuspy halos in our universe \cite{Colin,Bode, Avila-Reese}. Light neutralinos 
of some SUSY models can be CDM candidates, while sterile neutrinos \cite{Dodelson,AsakaBla,AsakaSha}, majorons \cite{Akhmedov,Berezinsky,Lattanzi} and light dark matter particles (LDM) \cite{BoehmFa} are proposed to be among the candidates for WDM. {While WDM can explain cuspy halos in our Universe,  it is ruled out by recent CMB observations and other data sets \cite{Madhavacheril}}. We refer in particular  to recent reviews \cite{Feng, Bergstrom} for the different dark matter models.\\

If dark matter consists of self-annihilating particles, one may expect to observe potential annihilation signals in regions where the dark matter density is particularly high, as in the center of our own galaxy. For instance, strong X-ray emission in the $511$~keV line from our galactic center can be interpreted as a sign of self-annihilating dark matter \cite{Jean, Weidenspointner, BoehmHo}. Other observations suggest an excess of GeV photons \cite{Boer2005}, microwave photons \cite{Hooper} and positrons \cite{Cirelli} in the galactic center. An interpretation in terms of self-annihilating dark matter is favored by the approximately isotropic emission of such particles and the absence of correlations with the Galactic disk.\\

Apart from the search for dark matter signals in our galactic center, models for their impact on various cosmological and astrophysical processes have been suggested in the literature. Both particle physics and cosmological data provide bounds on their annihilation cross section \cite{Bertone, Hutsi, Galli, Finkbeiner, Giesen, Evoli, Slatyer, Cline, Chulba} and potential signatures in the recombination spectrum have been explored by Chluba (2009). The annihilation of dark matter particles leads to the production of energy, part of which is injected into the IGM, where it can affect the ionization and thermal history of the Universe. In general the ionization of hydrogen and helium occurs through similar processes, and can therefore be described with a similar framework ({\it e.g.}\cite{Furlanetto2006}). Since dark matter particles are massive, their annihilation produces photons which have very high energies and typically provide minor contributions to the ionization of hydrogen and helium, unless extreme scenarios are considered \cite{Schleicher, Chuzhoy, Furlanetto2006}. On the other hand more energetic photons are required to ionize He$^+$ to He$^{++}$ and such photons are still rare among the first sources of light. Hence the contribution from the self-annihilation of dark matter is potentially more relevant in this case and will be explored in greater detail in this paper.\\

This paper is structured as follows. In \S\ref{1} we outline the different sources which affect the He$^{++}$ fraction. In subsection \S\ref{2}, we explain the ionization equation, and in subsection \S\ref{3}, we describe the contribution of quasars to the formation of He$^{++}$. In \S\ref{4} we discuss the effects of dark matter annhilation assuming a uniform dark matter density, while in \S\ref{5} we take into account the clumped dark matter density taking into account different halo
profiles. In sections \S\ref{6}, we describe our main results,  and discuss our conclusions in \S\ref{7}. The annhilation rates are evaluated in detail in {\bf the APPENDIX}

\section{Helium Reionization with quasars and dark matter}\label{1}

Singly ionized helium with a ionization threshold of $E_{\rm He^+,ion}=24.6$~eV is produced along with the
ionization of atomic hydrogen ($E_{\rm H^+, ion}=13.6$~eV), which is generally considered to occur as a result of
photoionization due to radiative feedback from the first stars and galaxies \cite{Barkana, Ciardi, Schleicher}. He$^+$ has a similar recombination rate as H$^+$, while the recombination rate is enhanced for He$^{++}$ by a factor of $5$.  In studies of helium reionization, it is thus typically assumed that the single-ionization of helium is more or less co-eval with hydrogen reionization, while the ionization of He$^{++}$ occurs at a later stage. The ionization of He$^+$ to He$^{++}$ further requires photons of higher energy owing to a greater ionization threshold ($E_{\rm He^{++}, ion}=54.4 eV$). Even in models considering reionization due to massive Pop.~III stars, the number of ionizing photons at $54.4$~eV is suppressed by at least a factor of $20$ compared to the peak of the emission. However, observational studies suggest that reionization is predominantly driven by low-luminosity galaxies at high redshift \cite{Bouwens}, implying a more conventional stellar population. As a result, the expected drop in the number of He$^+$-ionizing photons is even more significant, requiring the emission in quasars or potentially due to the self-annihilation of dark matter.

In the following subsections, we outline the framework to describe how helium is doubly-ionized from its singly ionized state due to photons produced
in quasars or as a result of dark matter annihilation. As shown in our calculation below, the formation of He$^{++}$ occurs during epochs where helium is already singly-ionized. We therefore adopt the single-ionization of helium as a general assumption in our calculation. While this could in principle lead to an overestimate of the He$^{++}$ production in epochs where the helium is still neutral, our results show that the latter happens to be inefficient during such times, therefore providing an a posteriori justification. For this study, we adopt a generic set of cosmological parameters approximately consistent with the recent PLANCK \cite{Planck} survey with $\Omega_{m} =0.31$ , $\Omega_{\Lambda}=0.68$, $\Omega_b =0.033$, $H=100h$ km $s^{-1} Mpc^{-1}$, $h=0.6711$, $n=0.96$, $\sigma_8 = 0.8$ and the critical density $\rho_{crit} = 3.1 \times 10^{17} h^{-1}~ M_\odot {\rm Mpc}^{-3}$. 

\subsection{IONIZATION EQUATIONS}\label{2}

While the main focus of this investigation concerns the epoch of helium reionization, a residual degree of doubly-ionized 
helium can be produced already in the epoch of recombination. In the absence of dark matter annihilation, it is well-known 
that helium recombines completely and becomes fully neutral. However, in the presence of dark matter annihilation, there will be an additional ionization term which prevents the He$^{++}$ abundances from recombining to zero, and instead a new equilibrium will be established corresponding to a non-zero doubly-ionized ionization degree. As already shown by Chluba (2009), helium is in the singly-ionized state until $z\sim2000$. Subsequently, a residual ionization of $\sim10^{-4}$ is maintained in the singly-ionized state. While the recombination rates of He$^{++}$ are higher by about a factor of 5 compared to He$^+$, residual abundances of up to $\sim10^{-5}$ can nevertheless be maintained. \\ 

As a minimal model  to obtain the residual ionization degree from the recombination epoch, we will in the following consider the ionization due to dark matter annihilation, the recombination to and the collisional ionization of He$^+$.  In addition to the processes outlined above, we consider the double-ionization of helium by energetic photons from high-redshift quasars. As we will show below, these are likely to give the dominant contribution leading to complete helium reionization in the intergalactic medium. The rate at which the mean ionization fraction of He$++$ evolves is then given by
\begin{eqnarray}
 \frac{d\bar{x}_{\rm He^{++}}}{dt}&=&k_{QSO}+k_{DM}+(1.-\bar{x}_{He^{++}}) n_e \beta_{He^{+}}\nonumber \\
 &-&\bar{C}\alpha_A (T)n_e\bar{x}_{He^{++}}. \label{eq1}
\end{eqnarray}
Here, the first term describes the ionization rate due to quasar photons, the second term describes the contribution of dark matter, the third term describes the collisional ionization by thermal electrons and the fourth term describes radiative recombination. The fraction of He$^{++}$  is expressed as $\bar{x}_{\rm He^{++}}=\frac{n_{He^{++}}}{n_{He}}$, while the number density of thermal electrons is $n_e$, and $\beta_{He^+}$ \cite{Cen} is the collisional ionization coefficient, which evolves with the matter temperature and is given as
\begin{eqnarray}
 \beta_{He^+} &=& 5.68 \times 10^{-12}T^{1/2}e^{-631515/T} \nonumber \\
 &\times&\left[1+\left(\frac{T}{10^5}\right)^{1/2}\right]^{-1} \mathrm{cm}^3 \mathrm{s}^{-1}.
\end{eqnarray}
The recombination coefficient $\alpha_A(T)$ \cite{Cen} is given as
\begin{eqnarray}
 \alpha_A(T)&=&3.36 \times 10^{-10}T^{-1/2}\left(\frac{T}{10^3}\right)^{-0.2}\nonumber \\
 &\times&\left[1+\left(\frac{T}{10^6}\right)^{0.7}\right]^{-1} \mathrm{cm}^3 \mathrm{s}^{-1}
\end{eqnarray}
and $\bar{C}=\langle \rho^2 \rangle / \langle \rho\rangle^2$ is the baryonic clumping factor which we describe as
\begin{equation}
 \bar{C}=
 \begin{cases}
  1 &\text{($z>15$)}\\
  1+\frac{15-z}{9} &\text{($6<z<15$)}\\
  3 &\text{($z<6$)} .
 \end{cases}
\end{equation}
In this context, we note in particular that a uniform baryon density is assumed at $z>15$, before a significant amount of structure formation has occured, and a generic value of $\bar{C}=3$ is adopted at $z<6$, as also employed in other studies (e.g.\cite{Furlanetto2008}). A linear interpolation is pursued in between. We note that the baryon clumping factor used here is smaller than in studies of hydrogen reionization \cite{Ciardi, Schleicher}, as helium reionization is strongly driven by rare objects like quasars and therefore proceeds on large spatial scales.\\ 

The processes outlined here are important at different cosmological epochs, and will lead to characteristic signatures of 
the evolution of He$^{++}$ at different cosmological times. In particular the evolution of the recombination and 
collisional ionization rate depends on the temperature of the gas. We have used the RECFAST code \cite{Seager} to follow the gas temperature up to $z=15$ and below $z=15$ we assume the IGM temperature rises to $T=10000$~K during cosmic reionization.

\subsection{CONTRIBUTION FROM QUASARS}\label{3}
Quasars and AGNs produce hard photons, which substantially contribute to the production of doubly-ionized helium. Following Furlanetto et. al. \cite{Furlanetto2008}, the doubly ionization rate of helium due to ionizing photons from quasars is given as
\begin{eqnarray}
k_{QSO}&=&\int dL_B \frac{\dot{N}}{\bar{n_{He}}}\frac{d\phi}{dL_{B}}  \label{eq2} \\
 \dot{N}&=&2\times 10^{55}\frac{L_B}{10^{12}L_{\odot}}~~~s^{-1}
\end{eqnarray}
where $\dot{N}$ is the total number of photons emitted per second by a quasar which have energy greater than the ionization threshold of He$^{++}$. Here $d\phi/dL_B$ is the quasar luminosity function (QLF), which is defined as the number of quasars per unit volume per unit luminosity interval. Following the work of Hopkins et. al. \cite{Hopkins}, we obtain the following expression:
\begin{equation}
 \frac{d\phi}{dlogL_B}=\frac{\phi_{*}}{(L/L_*)^{\gamma_1}+(L/L_*)^{\gamma_2}},
\end{equation}
where $\phi_{*}$ is the normalization, $L_{*}$ is the break luminosity, $\gamma_1$ is the faint-end slope and $\gamma_2$ is the bright-end slope.

\begin{figure*}
\begin{center}
\subfloat[]{ \includegraphics[height=3.5in,width=3.0in,angle=-90]{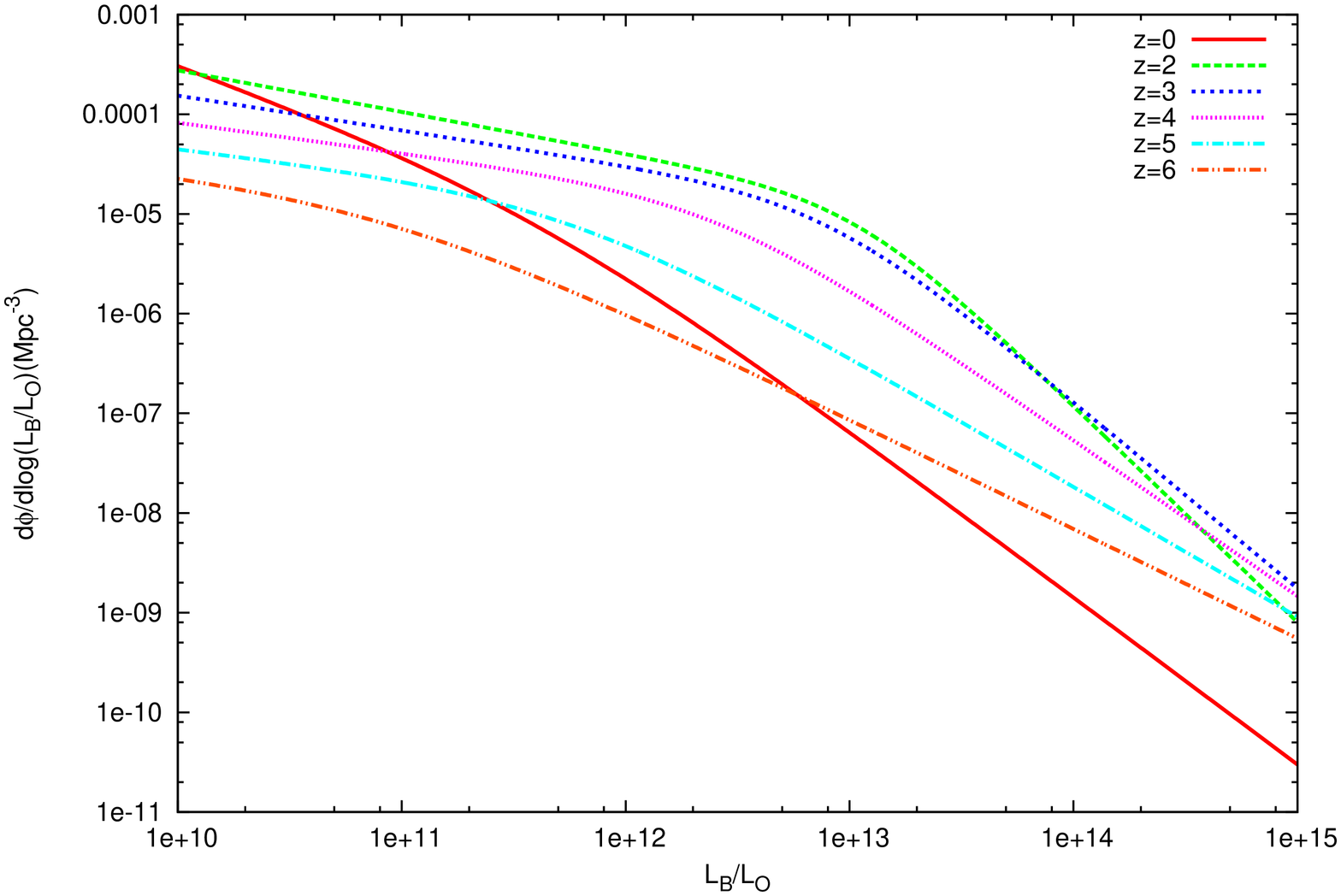}\label{gra1}}
\subfloat[]{ \includegraphics[height=3.5in,width=3.0in,angle=-90]{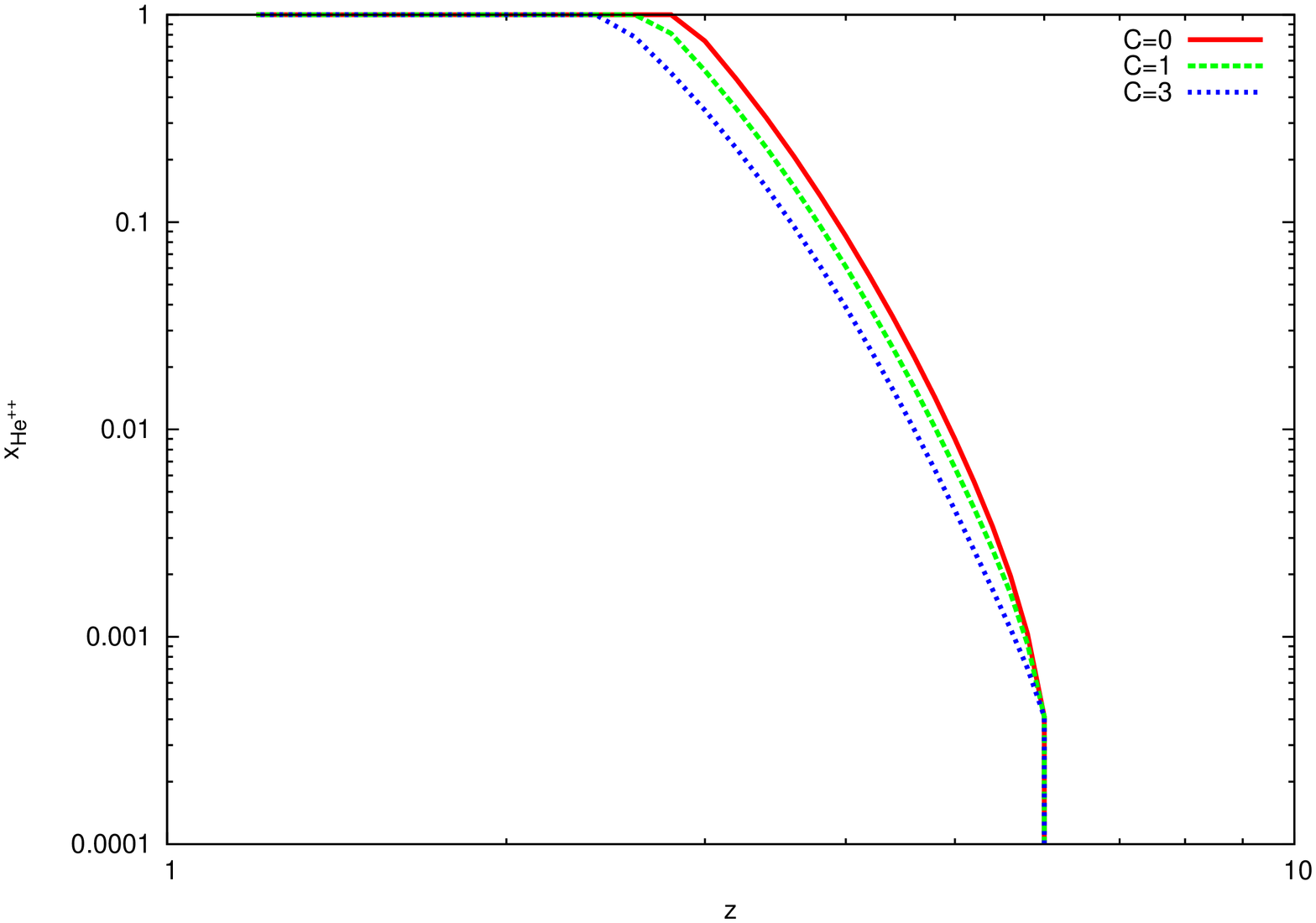}\label{gra2}}
 \caption{(a) The quasar luminosity function $\frac{d\phi}{dLog{L/L_{\odot}}}$ at different redshifts. \newline (b) Evolution of the He$^{++}$ fraction in the presence of quasars without dark matter annihilation.} 
 \end{center}
 \end{figure*} 

The equation for the QLF corresponds to a double power law which is also evident from Fig.[\ref{gra1}]. The parameters involved in the expression for the QLF evolve as 
\begin{eqnarray}
 log~L_{*}&=&(log~L_*)_0+k_{L1}\xi+k_{L2}\xi^2+k_{L3}\xi^3, \nonumber\\
 \gamma_1 &=& (\gamma_1)_0 \times 10^{k_{\gamma_1} \xi}, \nonumber\\
 \gamma_2 &=& \frac{2(\gamma_2)_0}{10^{k_{\gamma_{2,1}}\xi}+10^{k_{\gamma_{2,2}}\xi}}, \nonumber \\
 \xi &=& log \left(\frac{1+z}{1+z_f}\right)
\end{eqnarray}
where $k_{L1}$, $k_{L2}$ and $k_{L3}$ are free parameters and $z_{f}=2$. The best fit parameters for the above model are given in [Table \ref{table1}]\cite{Hopkins},
\begin{table}[ht]
 \caption{Best fit Parameters for QLFs \cite{Hopkins}}
 \centering
 \begin{tabular}{| c | c |}
 \hline\hline
 Parameters & Values \\[0.5ex]
 \hline
 $log \phi_{*}$ & $-4.825\pm 0.060$ \\
 $log L_{*}$ & $13.036 \pm 0.043$  \\
 $k_{L1}$ & $0.632 \pm 0.077$ \\
 $k_{L2}$ & $-11.76 \pm 0.38$  \\
 $k_{L3}$ & $-14.25 \pm 0.80$  \\
 $(\gamma_1)_0$ & $0.417 \pm 0.055$ \\
 $k_{\gamma_1}$ & $-0.623 \pm 0.132$ \\
 $(\gamma_2)_0$ & $2.174 \pm 0.055$ \\
 $k_{\gamma_{2,1}}$ & $1.460 \pm 0.096$  \\
 $k_{\gamma_{2,2}}$ & $-0.793 \pm 0.057$ \\[1ex]
 \hline
 \end{tabular}
\label{table1}
\end{table}

where $\phi_{*}$ is in units of Mpc$^{-3}$ and $L_{*}$ is expressed in terms of $L_{\odot}$. The effect of ionization by quasars has been included from $z\sim6$ since that is the epoch where the first quasars have been observed, and from which we expect the adopted quasar luminosity function to be valid. The ionizing photons from quasars lead to a rapid double-ionization of helium. The timescale for this process is much smaller than the timescale for collisional ionization. Figure~[\ref{gra1}] shows the QLF as a function of luminosity for different redshifts. The impact of quasars on helium reionization is given in Figure~[\ref{gra2}], which shows the evolution of the He$^{++}$ fraction as a function of redshift assuming different clumping factors $\bar{C}$ for the baryons. Due to the effect of recombination, the amount of He$^{++}$ decreases with the baryonic clumping factor. While the calculation here implicitly assumes that all quasar spectra are the same, observational investigations have shown that characteristic quasar spectra can be obtained when considering a larger sample of sources, so that a well-defined statistical average exists \cite{Sazonov04, Sazonov08}. We expect that the latter will be sufficient to describe the large-scale average of helium reionization. So far, even the highest-redshift quasar known to date at z=7.085 still shows the same characteristic features as quasars at lower redshift \cite{Mortlock}, therefore providing no evidence for a cosmological evolution. If such an evolution is found in the future, it could further modify the results for the cosmic reionization of helium obtained here.

\subsection{EFFECTS OF DARK MATTER}\label{4}
Dark matter constitutes about $28\%$ of our Universe \cite{Planck}. While dark matter candidates need to be extremely stable with long lifetimes and small annihilation cross sections, the annihilation of even a small fraction of the total abundance can have a significant effect on the reionization of hydrogen and helium. We adopt here the typical framework employed to consider the impact of dark matter on hydrogen reionization \cite{Schleicher}, and adapt it for helium. In case of a uniform dark matter distribution, we obtain
\begin{eqnarray}
&k_{DM}=\eta_2\chi_i\langle \sigma v \rangle\left(\frac{m_p c^2}{E_{\rm He^{+}, ion}}\right)&\nonumber\\
&\times\frac{\Omega_{\chi}\rho_{crit}}{m_{\chi}} \left(\frac{\Omega_{\chi}}{\Omega_b}\right)(1+z)^3& ,\label{eq3}
 \end{eqnarray}
 where $\langle\sigma v \rangle=3 \times 10^{-26} cm^3s^{-1}$  is the thermally averaged cross section for annihilation,  $\chi_i$ is the fraction of energy that goes into ionizing He$^{+}$ and is defined as $\chi_i=E/E_{ph}$ where $E$ is the energy that goes into ionization while $E_{ph}$ is the energy of the incoming particle. The total energy required for ionization is $E=N \times E_{ion}$ where $N$ is the total number of ionization and $E_{ion}$ is the ionization energy. From \cite{Dalgarno} we have the mean energy per ion pair $W$ ({\it i.e.} $W=E_{ph}/N$) which is parametrically given as $W=W_0(1+Cx^{\alpha})$. Thus substituting, we have $\chi_i=E_{ion}/W$. For photons with energies greater than $1$~keV and a cosmological mixture of  hydrogen and helium gases the value of the constant parameters are $W_0=16400$~eV, $C=11.7$ and $\alpha=1.05$. For He$^+$ ionization we assume $\eta_2 = 8/0.24$, while the ionization energy $E_{\rm He^{++}, ion}=54.4$~eV. 
 
Since the dark matter annihilation rate depends on the dark matter density, we start our calculation at $z=1000$ to account for the formation of the residual amount of He$^{++}$ from the annihilation at early stages. We will in the following consider both warm dark matter particles with masses of about $20$~MeV, and cold dark matter particles with masses of about $80$~GeV.

\subsection{DARK MATTER CLUMPING}\label{5}

In the previous section we have assumed that the dark matter is uniformly distributed. While this uniform density of dark matter can be a good approximation at early stages, it is expected to clump considerably during structure formation. This non-uniform density distribution can be accounted for with a clumping factor for the dark matter, which effectively describes how the dark matter annihilation is enhanced due to the dark matter clumping or by methods as mentioned in \cite{Hutsi}. The latter depends both on the abundance of dark matter halos, as well as their characteristic density profiles.\\ 

Here, we consider the Moore, NFW and Burkert profiles for the dark matter \cite{Burkert,Moore,NFW1996,NFW1997,Salucci} and adopt the approach by Cumberbatch et al. \cite{Cumberbatch} to calculate the clumping factor. We first consider the total annihilation rate resulting from dark matter halos with a minimum mass $M_{min}$ and a maximum mass $M_{max}$ at a redshift $z$ which is given as

\begin{figure*}[htb!]
\begin{center}
\includegraphics[height=4.5in,width=2.5in,angle=-90]{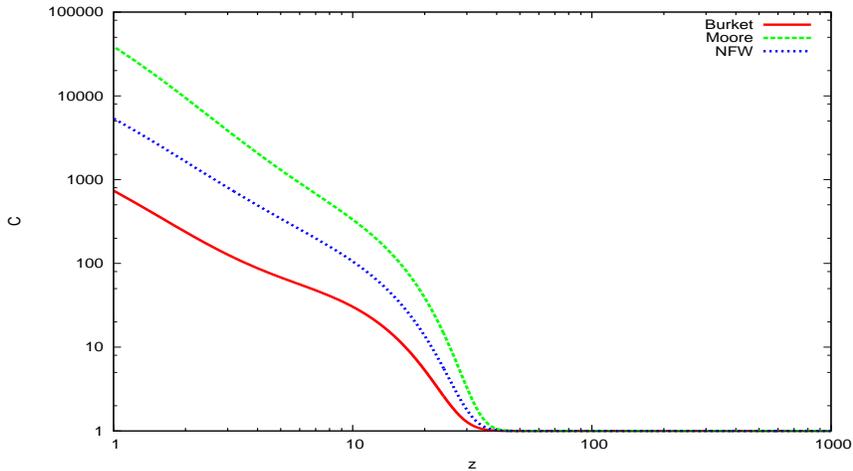}  
 \caption{Evolution of the dark matter clumping factor for the NFW, Moore and Burkert dark matter density profiles.}\label{gra3}
\end{center}
 \end{figure*}
\begin{figure*}[htb!]
\begin{center}
\subfloat[]{\includegraphics[height=3.2in,width=2.6in,angle=-90]{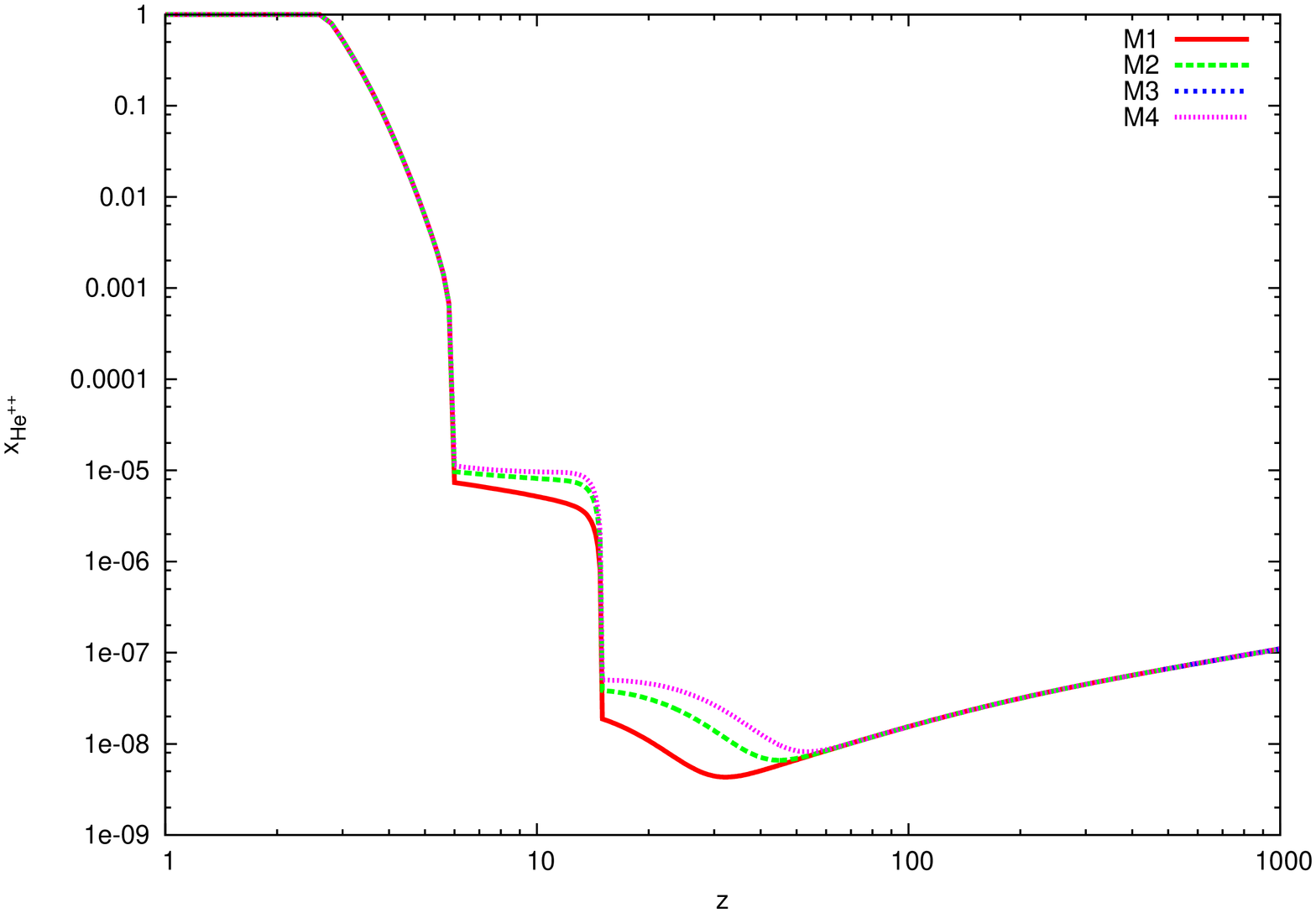} \label{gra11}} \\
\subfloat[]{\includegraphics[height=3.2in,width=2.6in,angle=-90]{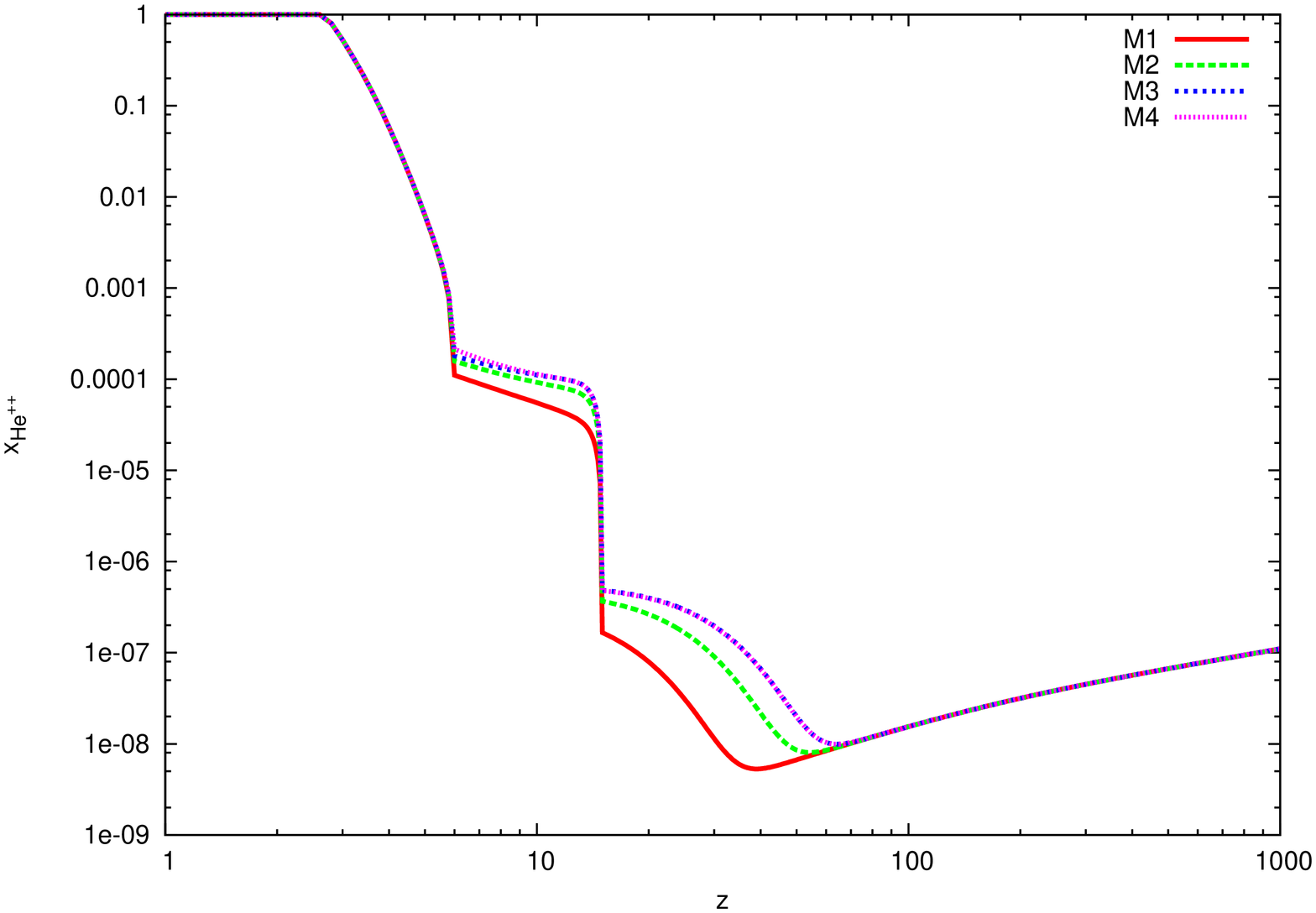}  \label{gra12}}
\subfloat[]{\includegraphics[height=3.2in,width=2.6in,angle=-90]{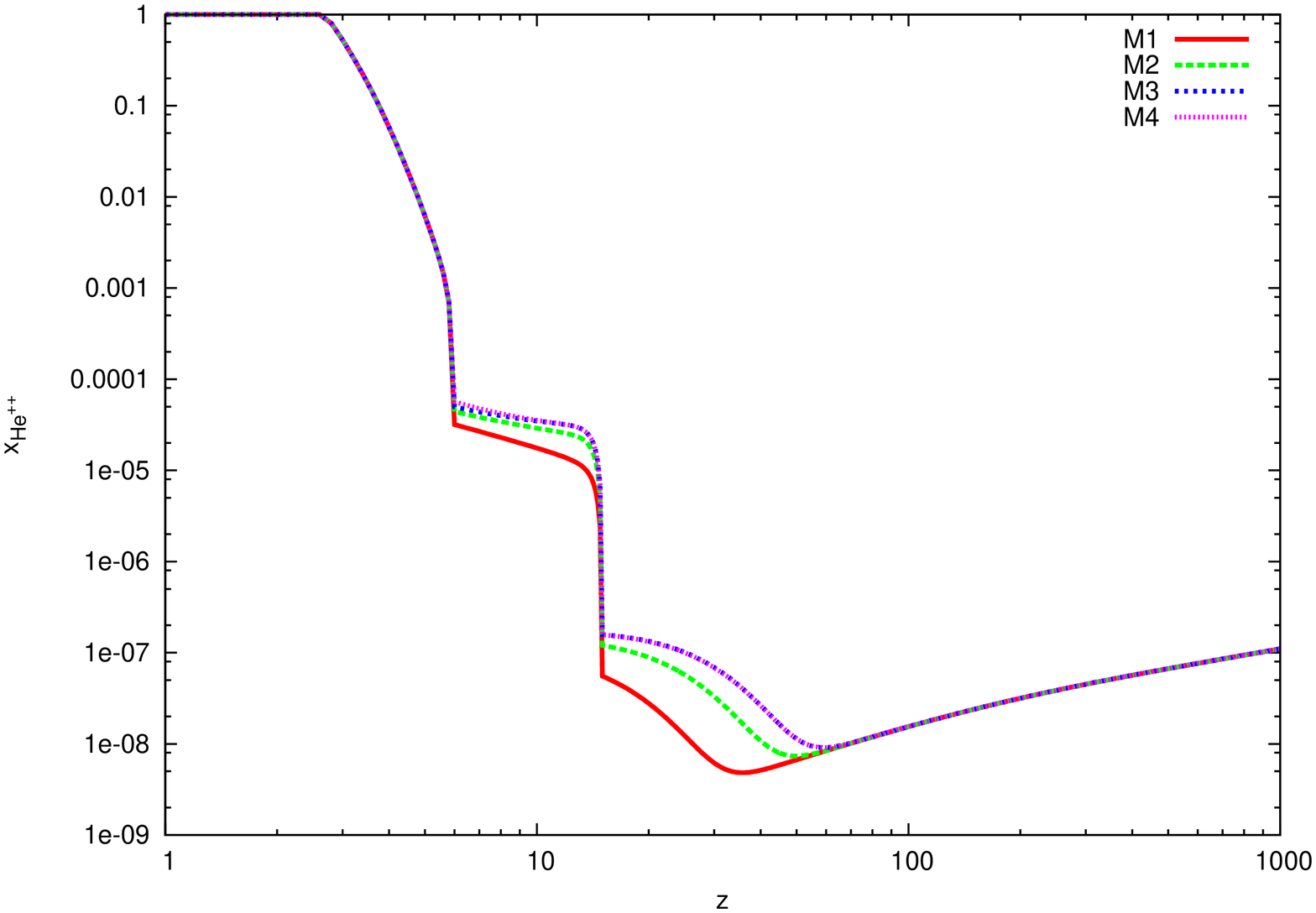} \label{gra13}}
  \caption{Evolution of the He$^{++}$ fraction for cold dark matter assuming [a] a Burkert profile, [b] a Moore profile and [c] an NFW profile. We show that the results depend only weakly on the mass range considered, where $M_1 \rightarrow 10^6-10^{12}$~M$_\odot$, $M_2 \rightarrow 10^4-10^{12}$~M$_\odot$, $M_3 \rightarrow 10^{-12}-10^6$~M$_\odot$ and $M_4 \rightarrow 10^{-12}-10^6$~M$_\odot$ with $M_{cut}=10^6 M_{\odot}$} \label{cap3}
\end{center}
\end{figure*}

 
\begin{eqnarray}
 \Gamma_{halos}(z)&=&(1+z)^3 \nonumber \\
 &\times&\int_{M_{min}}^{M_{max}}dM\frac{dn}{dM}(M,z)R(M,z),
\end{eqnarray}
where $dn/dM$ is the mass function which gives the number of halos with mass $M$ and $R(M,z)$ is the rate of annihilation inside a halo of mass $M$ at the redshift $z$. As discussed in the appendix, we will typically adopt a lower limit of $10^6$~M$_\odot$ for the halo masses, even though it can be as low as $10^{-12}$~M$_\odot$ in WDM models \cite{Cumberbatch}. As shown in Fig.~\ref{cap3}, even significant changes in this parameter do not strongly affect our results. 

We evaluate the halo mass function using the Press-Schechter formalism \cite{Press}. The annihilation rate inside a halo of mass $M$ at redshift $z$ is then
\begin{equation}
 R(M,z) = \frac{1}{2}\frac{\langle \sigma_{\chi}v \rangle}{m_{\chi}^2}\int_{r=0}^{r_{vir(M,z)}}\rho^2(r)4\pi r^2dr
\end{equation}
where $r_{vir}$ is the virial radius. The annihilation rate in the smooth background is 
\begin{equation}
 \Gamma_{smooth}(z)=\frac{1}{2}\frac{\langle \sigma_{\chi}v \rangle}{m_{\chi}^2}\bar{\rho}^2_{\chi}(z)
\end{equation}
and thus the clumping factor at a given redshift(z) is given as
 
\begin{equation}
 C_{halo}(z)=1+\frac{\Gamma_{halos}(z)}{\Gamma_{smooth}(z)}.
\end{equation}

The annihilation rate inside the halos can now further be determined for different dark matter density profiles. In general, the density distribution can be described as a continous function of $r$ \cite{Cumberbatch}:
\begin{equation}
 \rho(r)=\frac{\rho_s}{(r/r_s)^{\gamma}[1+(r/r_s)^{\alpha}]^{(\beta-\gamma)/\alpha}}. \label{eq5}
\end{equation}
\begin{figure*}[htb!]
\subfloat[]{ \includegraphics[height=2.2in,width=2.5in,angle=-90]{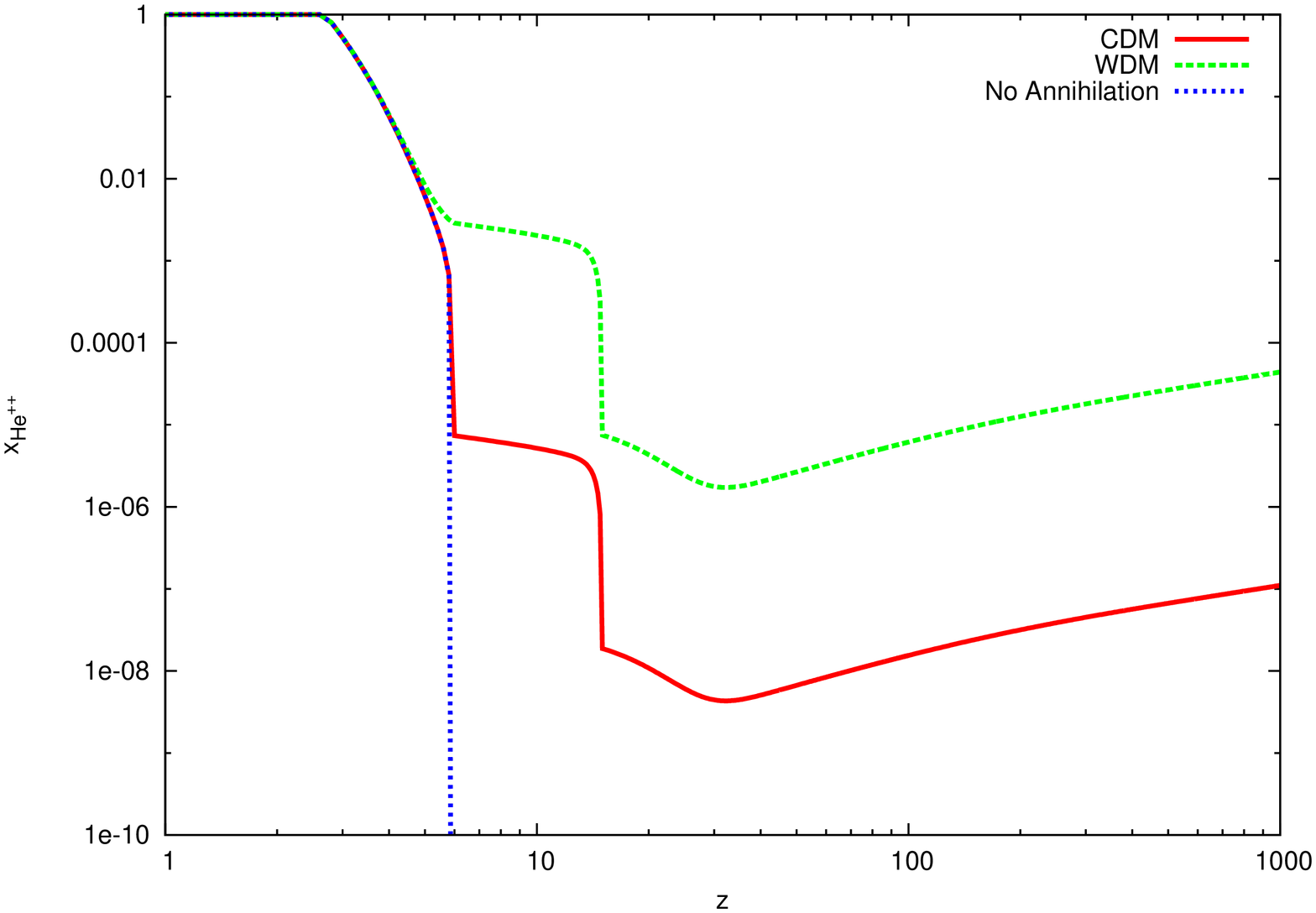} \label{gra4}}
\subfloat[]{ \includegraphics[height=2.2in,width=2.5in,angle=-90]{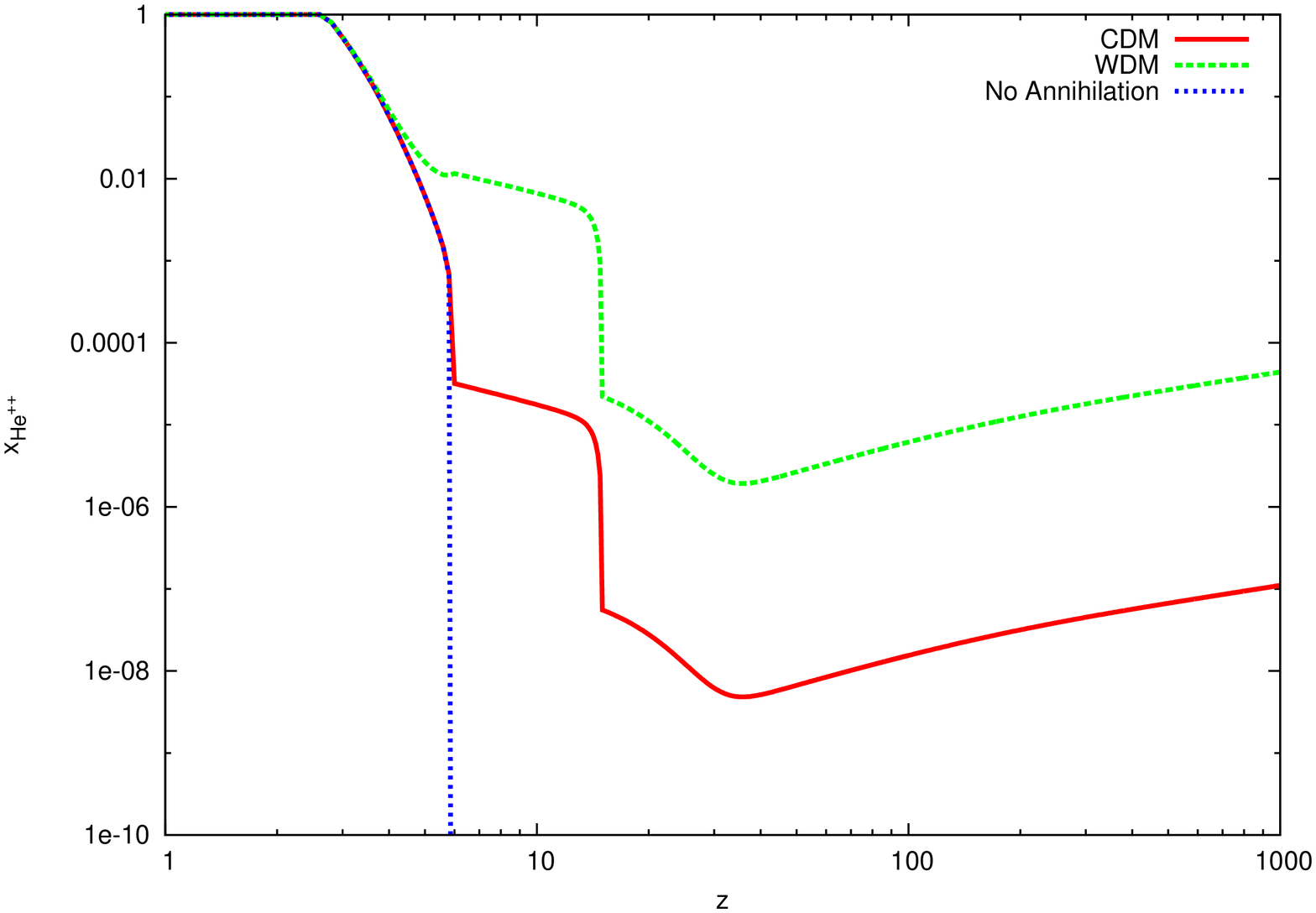}  \label{gra5}}
\subfloat[]{ \includegraphics[height=2.2in,width=2.5in,angle=-90]{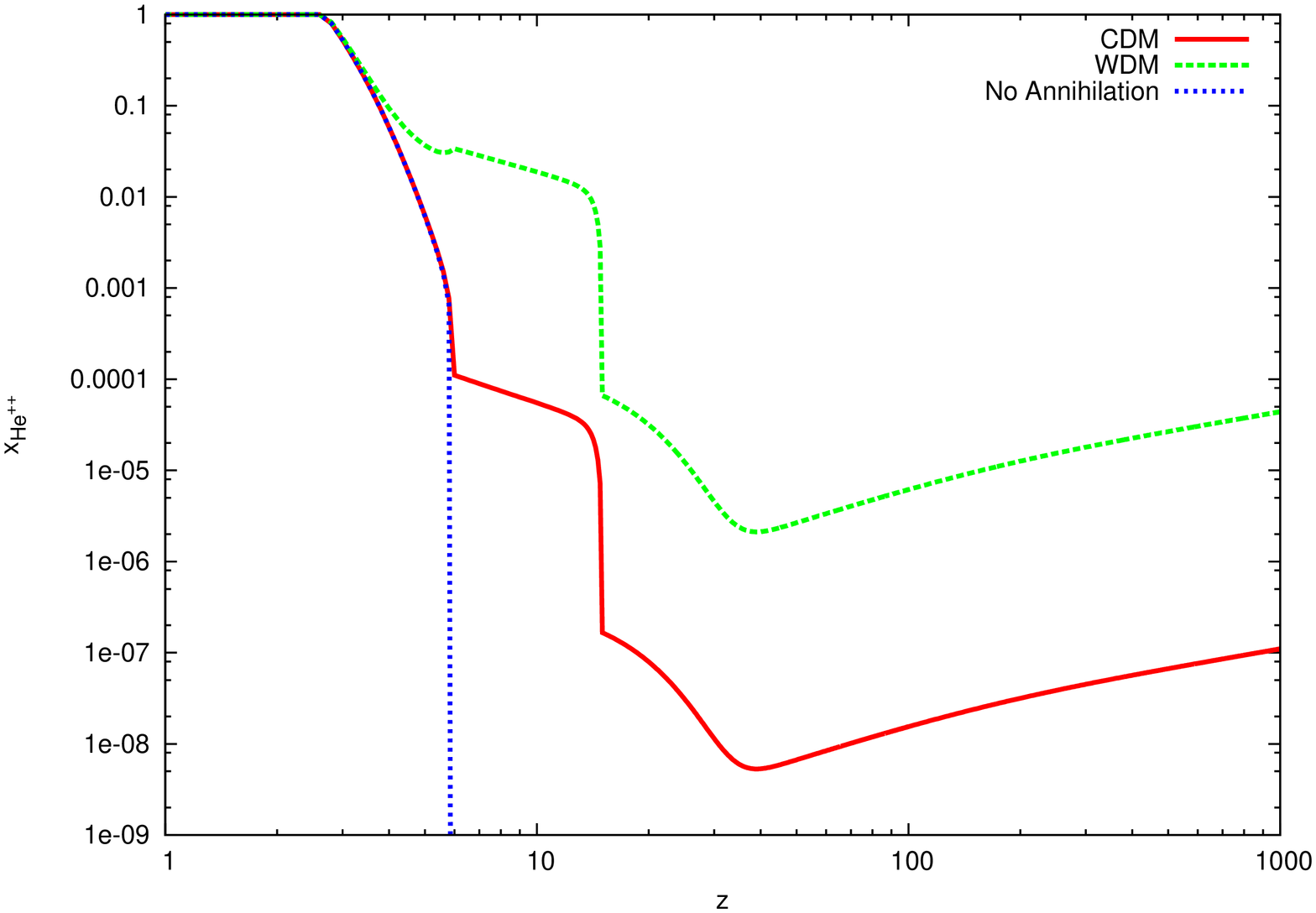} \label{gra6}}
 \begin{center}
\caption{Evolution of He$^{++}$ in the presence of quasars and dark matter annihilation for a warm and cold dark matter scenario and compared with the case when there is no dark matter annihilation. [\ref{gra4}] Burkert profile, [\ref{gra5}] NFW profile and [\ref{gra6}] Moore profile.} \label{cap1}
\end{center}
\end{figure*}
We will focus in particular on the NFW, Moore and Burkert profile \cite{Burkert,Moore,NFW1996,NFW1997,Salucci}. The derivation of the rates of annihilation for each of the profiles is given in the Appendix.
\begin{figure*}[htb!]
\begin{center}
\includegraphics[height=4.5in,angle=-90]{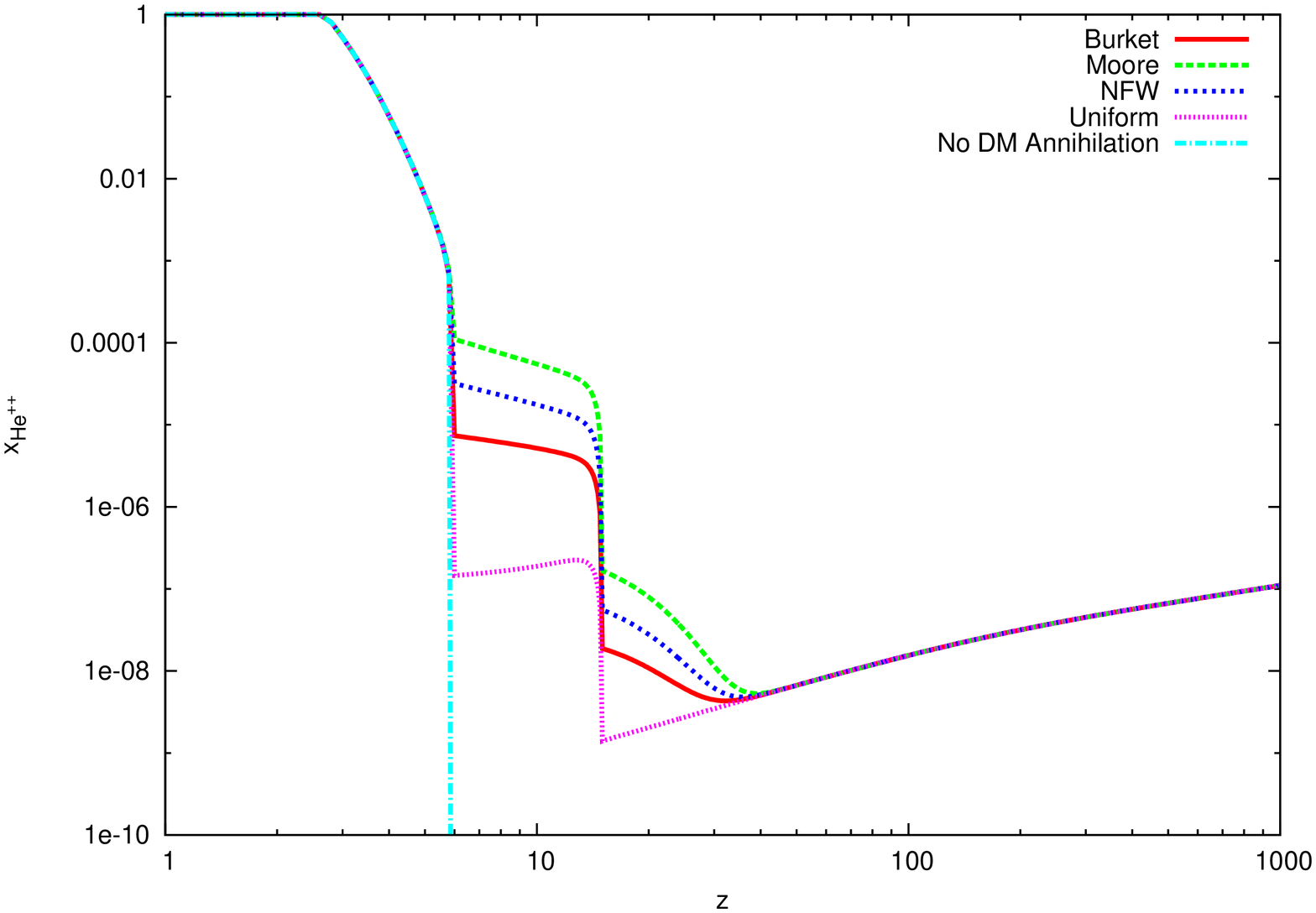} \label{gra7}
 \caption{Evolution of the He$^{++}$ fraction for a cold dark matter scenario in case of a uniform distribution, and for the case of clumpy dark matter assuming Moore, NFW and Burkert profiles. The mass range of the dark matter halos is considered to range from $10^6$~M$_\odot$ to $10^{12}$~M$_\odot$. It also compares the ionization fraction when no dark matter annihilation takes place.} \label{cap2}
\end{center}
\end{figure*}

The ionization rate from dark matter annihilation [\ref{eq3}] is now modified to include the clumping factor and can be written as
\begin{eqnarray}
&k_{DM}=\eta_2\chi_i C_{halo}(z)\langle \sigma v \rangle \nonumber \\
&\times\left(\frac{m_p c^2}{E_{ion}}\right)\frac{\Omega_{\chi}\rho_{crit}}{m_{\chi}} \left(\frac{\Omega_{\chi}}
{\Omega_b}\right)(1+z)^3.\label{eq4}
\end{eqnarray}

\section{Results}\label{6}

The evolution of helium reionization in the presence of quasars and in the absence of dark matter annihilation is given in Fig.~[\ref{gra2}] for baryonic clumping factors ranging from $0$ (underdense regions with inefficient recombination) to $3$ (overdense regions with enhanced recombination). The results show a very steep increase of the He$^{++}$ fraction as a function of redshift, with helium reionization being completed at $z\sim2.5$. The latter reflects both the shift of the quasar luminosity function towards higher luminosities, as well as the increase of the quasar abundances at a given luminosity in Fig.~[\ref{gra1}]. The results depend only weakly on the baryonic clumping factor in the calculation.\

The rate of dark matter annihilation depends on the number density of the dark matter particles given by $n_{\chi}=\rho_{\rm dm}/m_{\chi}$, where $\rho_{\rm dm}=\Omega_{dm}\rho_{\rm crit}$, with $\rho_{\rm crit}$ the critical mass density and  $\Omega_{dm}\sim28\%$. The dark matter content of our Universe may consist of massive cold dark matter particles $(m_{\chi}>10 GeV)$ or warm dark matter with characteristic masses $m_{\chi}\sim$~MeV. Due to the large difference in the masses of warm and cold dark matter, there is a corresponding difference in the number density of dark matter particles for these two models. In Fig.~[\ref{cap1}], we compare the results when considering the contributions of cold and warm dark matter to helium reionization, finding that the contribution from warm dark matter is clearly enhanced due to the higher annihilation rate. However, the warm dark matter scenario is ruled out by the PLANCK survey \cite{Madhavacheril}, so we will subsequently focus on the cold dark matter scenario. \

We can generally distinguish a number of characteristic phases in the evolution of He$^{++}$: At $z<1000$, the He$^{++}$ abundance decreases but remains non-zero, reflecting the decrease of the annihilation rate and the decrease in the gas temperature, which is initially coupled to the temperature of the CMB.  At $z\sim30$, the effects of dark matter clumping become noticable which starts increasing the dark matter annihilation rate and hence the He$^{++}$ fraction, which is further enhanced when the gas temperature increases to $\sim10^4$~K due to hydrogen reionization. Subsequently, we find a further moderate increase until the contribution from quasars kicks in at $z=6$.\

We notice that these phases can be identified both in the calculations with cold and warm dark matter and for all the halo profiles considered, including the NFW, Burkert and Moore profile. The evolution of the dark matter clumping factor is given for these cases in Fig.~[\ref{gra3}]. As expected, we find the highest clumping factor for the Moore profile, which is the steepest density profile, while it is lowest for the Burkert profile, in agreement with the study by Cumberbatch et al. \cite{Cumberbatch}. We expect that the real dark matter clumping factor will lie in between these extreme cases. \

A comparison of the implications of the different clumping factors for the cold dark matter scenario including also a case with a uniform dark matter distribution is shown in Fig.~[\ref{cap2}]. It is clearly visible that the He$^{++}$ abundance is reduced by at least a few orders of magnitude in case of a uniform DM density. Apart from the increase in the He$^{++}$ fraction due to the temperature increase at $z\sim15$, the evolution of the ionized fraction is generally decreasing with decreasing redshift at $z>6$, reflecting the decreasing density of dark matter and the decreasing annihilation rate. This behavior is inverted when clumpy dark matter is considered, as the clumpiness increases with redshift and compensates for the decrease of the mean dark matter density. As a result, the contribution of dark matter annihilation may rise rapidly at $z<6$, when also the abundance of luminous quasars is rapidly increasing. At $z<6$, the fraction of He$^{++}$ increases both due to the increased temperature of the IGM, as well as due to the increasing dark matter clumping factor, which varies between $\sim100$ and $\sim3000$ at $z=6$ for the scenarios considered here. As a result, the He$^{++}$ abundance is clearly enhanced for clumpy dark matter compared to a uniform distribution, and thus rises more steeply. A close zoom-in shows that the rise is even steeper for the Moore profile compared to NFW or Burkert, but then converges from $z\sim4$ when the abundance from quasars becomes important. The completion of helium reionization may thus change slightly by $\Delta z\sim0.5$ due to the additional contribution, even though the latter is hard to distinguish from the expected evolution in the presence of quasars. As the formation of quasars is expected to follow the build-up of structures, we expect that both contributions may evolve in a similar fashion, thus preventing a significant change in the evolution at late times. \

Finally, we also checked how the results depend on the adopted integration range when we determine the dark matter clumping factor. As visible in Fig.~[\ref{cap3}], there are mostly minor changes by at most one order of magnitude, even in cases where the integration range is changed by several orders of magnitude. Even though there is a physical uncertainty on the appropriate range to consider, we find that the overall behavior is rather robust and we can again identify the same characteristic phases in these calculations.

\section{Discussion and Conclusion}\label{7}
We have investigated the impact of self-annihilating clumpy dark matter on the evolution of the He$^{++}$ abundance over cosmic history. We find that the contribution of dark matter is potentially relevant before quasar feedback becomes important, and provides a residual He$^{++}$ fraction already at high redshift that depends on the dark matter model. \ 

In models with a uniform dark matter distribution, the contribution from dark matter annihilation decreases with decreasing redshift due to the decreasing dark matter densities, which is generally reflected in decreasing abundances of He$^{++}$ before the feedback from quasars becomes more relevant. In models of clumpy dark matter, the clumpiness however increases with time as a result of structure formation, and therefore both the annihilation rate and the He$^{++}$ abundance increases with decreasing redshift. We further note that the thermal evolution of the IGM is relevant for the residual abundance of He$^{++}$, as the increased temperatures due to hydrogen reionization give rise to reduced He$^{++}$ recombination rates and therefore a higher net abundance. A similar increase occurs at $z<6$ when the temperature of the IGM raises to $\sim20000$~K. \

While in the case of a uniform dark matter distribution, the annihilation rate eventually becomes negligible at low redshift, there is a considerable enhancement in the presence of clumpy dark matter, corresponding to a clumping factor of $10^2-3\times10^3$ at $z=6$ and a clumping factor of $3\times10^3-3\times10^5$ at $z=1$, which outweighs the decrease of the dark matter density with redshift. As a result, there can be a substantial contribution at $z\sim3$ where helium reionization is expected to approach completion. The effect is however highly degenerate with the contribution from quasars, as both their abundance and peak luminosity is expected to increase towards $z\sim2-3$, reflecting the formation and further growth of cosmological structures. We therefore expect that such a contribution may be present, even though it should be difficult to infer how much the completion of helium reionization has changed as a result. \

To determine such potential contributions, it is thus necessary to both accurately map out the evolution of helium reionization, extending current observational studies \cite{Jakobsen, Heap, Davidsen, Fechner, Reimers1997, Kriss, Smette, Shull, Anderson, ZhengKr, ZhengMe, Reimers2005}, but also to pursue a precise number count of X-ray sources throughout the evolution of the Universe, including both Compton-thin \citep[][]{Shankar} and Compton-thick \cite{Treister} sources. \

Rather than considering the completion of helium reionization, it may be helpful to probe the cosmic evolution of He$^{++}$ at earlier stages by extending current studies of the He$^+$ Lyman $\alpha$ forest and comparing the latter with the characteristic evolution of the quasar population over time. It is at least conceivable that the slope of the evolution provides additional information, which will be valuable to distinguish different scenarios. A further assessment of the uncertainties in the observed population and its implications for the available amount of X-rays for helium reionization will be necessary in this respect.

We note here that our calculations adopted a mass of 80 GeV for the dark matter particles, along with a cross-section of $\langle \sigma v\rangle = 3\times10^{-26}$~cm$^3$~s$^{-1}$ as consistent with the Planck data \cite{Planck}. We note that for larger cross sections, the annihilation rate would be enhanced, and the contribution to helium reionization thus more significant, while the effect becomes less important for lower cross sections. Similarly, the mass of the dark matter particles determines their number density. In such a case, we still expect the same  effects to occur, as the dark matter effects considered here depend only on the ratio $\langle \sigma v\rangle/m_{\chi}$

\section*{ACKNOWLEDGEMENTS}
BB is grateful to the German Academic Exchange Service (DAAD) for the funding of a collaborative visit via the program ''A New Passage to India'', and her supervisor Prof. T. R. Seshadri for the support of this project. DRGS is indebted to the Scuola Normale Superiore in Pisa for an invitation as a ''Distinguished Scientist'' in November 2014, where part of this manuscript has been created.

\section*{APPENDIX}
\appendix 
\label{APPENDIX}
\numberwithin{equation}{subsection}
\subsection{NFW PROFILE}
The NFW profile is a special case of eqn.(\ref{eq5}) with the values $\alpha=1$, $\beta=3$ and $\gamma=1$. The rate of annihilation $R(M,z)$ inside such a halo can be calculated to give
\begin{eqnarray}
 R(M,z)&=&\frac{1}{2}\frac{\langle \sigma_{\chi}v \rangle}{m_{\chi}^2}\rho_s^2\frac{4\pi}{3}\left(\frac{r_{vir}(z,M)}{c_{vir}(z,M)}\right)^3\nonumber \\
 &\times&\left[1-\frac{1}{[1+c_{vir}(M,z)]^3}\right]
\end{eqnarray}
where 
\begin{eqnarray}
 &\rho_s(M,z)=\frac{M}{4\pi \left(\frac{r_{vir}}{c_{vir}}\right)^3}\nonumber \\
 &\times\frac{1}{\left[log[1+c_{vir}]-\left(\frac{c_{vir}}{[1+c_{vir}]}\right)\right]}. \nonumber
\end{eqnarray}

\subsection{MOORE PROFILE}
For the Moore profile, we have $\alpha=1.5$, $\beta=3$ and $\gamma=1.5$. The annihilation rate inside such a halo is calculated as
\begin{eqnarray}
 &R(M,z)=\frac{1}{2}\frac{\langle \sigma_{\chi}v \rangle}{m_{\chi}^2}\rho_s^2\frac{4\pi}{3}\left(\frac{r_{vir}(z,M)}{c_{vir}(z,M)}\right)^3\nonumber \\
 &\times F_1(c_{vir},x_{min})
\end{eqnarray}
where
\begin{eqnarray}
 &F_1(c_{vir},x_{min})=\frac{1}{3}\frac{1}{(1+x_{min})^2}\nonumber \\
 &+\frac{1}{1.5}\left[log\left(\frac{c_{vir}^{1.5}(1+x_{min}^{1.5})}{x_{min}^{1.5}(1+c_{vir}^{1.5})}\right)\right]\nonumber \\
 &+\frac{1}{1.5}\left[\frac{1}{1+c_{vir}^{1.5}}-\frac{1}{1+x_{min}^{1.5}}\right]
\end{eqnarray}
and
\begin{eqnarray}
 &\rho_s(M,z)=\frac{M}{4\pi\left(\frac{r_{vir}}{c_{vir}}\right)^3}\nonumber \\
 &\times\left[\frac{1}{1.5}log\left(\frac{1+c_{vir}^{1.5}}{1+x_{min}^{1.5}}\right)+\frac{1}{3}\frac{x_{min}^{1.5}}{1+x_{min}^{1.5}}\right]^{-1}\nonumber
\end{eqnarray}
where $x_{\rm min}=10^{-8}$.

\subsection{BURKERT PROFILE}
Apart from the NFW or Moore profiles, one often considers the Burkert density profile, which can also be written as a continous function of the form
\begin{equation}
 \rho(r)=\frac{\rho_s}{[1+(r/r_s)][1+(r/r_s)^2]}.
\end{equation}
For such a profile, the annihilation rate follows as
\begin{eqnarray}
&R(M,z)=\frac{1}{2}\frac{\langle \sigma_{\chi}v \rangle}{m_{\chi}^2}\rho_s^2 2\pi \left(\frac{r_{vir}(z,M)}{c_{vir}(z,M)}\right)^3\times \nonumber\\
 &\left[1-\frac{1}{2(1+c_{vir})}-\frac{1}{2}tan^{-1}(c_{vir})-\frac{1}{2(1+c_{vir}^2)}\right] 
\end{eqnarray}
where
\begin{eqnarray}
&\rho_s(M,z)=\frac{M}{4\pi\left(\frac{r_{vir}}{c_{vir}}\right)^3}\times\nonumber \\
&\left[log(1+c_{vir})+\frac{1}{2}log(1+c_{vir}^2)-tan^{-1}(c_{vir})\right]^{-1}. 
\end{eqnarray}

In the above expressions $c_{vir}$ is the concentration parameter and is defined as $c_{vir}(M,z)=K\frac{1+z_c}{1+z}$. We can evaluate the collapse redshift $z_c$ with the condition $(1+z_c)=\frac{\delta_{sc}(z)}{\delta_{sc}(z=0)}$ where
$\delta_{sc}(z)=\sigma(M_{*}(z))$ with $M_{*}=0.015~M$. The virial radius $r_{vir}$ is defined as $r_{vir}=\left(\frac{M}
{\frac{4\pi}{3}200\rho_m(z)}\right)^{\frac{1}{3}}$. In order to calculate the value of $M_{*}$ for very small halos
 we choose a cutoff mass $M_{cut}$ below which the concentration parameter remains invariant. As \cite{Cumberbatch}, we adopt a typical mass scale of  $10^6 M_{\odot}$, corresponding to the mass of the halos where the first stars are expected to form \cite{Abel, Bromm, Latif}. To evaluate the mass integral, the mass limits have been chosen to be $M_{min}=10^6 M_{\odot}$ and
$M_{max}=10^{12} M_{\odot}$. We have checked that our results are not sensitive to this choice.\\


\end{document}